\documentstyle[preprint,aps]{revtex}

\begin{document}
\title{Reflection High-Energy Electron Diffraction oscillations during epitaxial
growth of artificially layered films of $(BaCuO_{x})_{m}/(CaCuO_{2})_{n}$}
\author{G. Balestrino, S. Lavanga, S. Martellucci, P. G. Medaglia, A. Paoletti, G.
Pasquini, G. Petrocelli, A. Tebano, A. Tucciarone}
\address{INFM-Dipartimento di Scienze e Tecnologie Fisiche ed Energetiche, Universit%
\`{a} di Roma ''Tor Vergata'', Via di Tor Vergata 110, 00133 Roma.}
\date{\today}
\maketitle

\begin{abstract}
Pulsed Laser Deposition in molecular-beam epitaxy environment (Laser-MBE)
has been used to grow high quality $BaCuO_{x}/CaCuO_{2}$ superlattices. In
situ Reflection High Energy Electron Diffraction (RHEED) shows that the
growth mechanism is 2-dimensional. Furthermore, weak but reproducible RHEED
intensity oscillations have been monitored during the growth. Ex-situ x-ray
diffraction spectra confirmed the growth rate deduced from RHEED
oscillations. Such results demonstrate that RHEED oscillations can be used,
even for $(BaCuO_{x})_{2}/(CaCuO_{2})_{2}$ superlattices, for phase locking
of the growth.
\end{abstract}

\bigskip 

In 1981, the first observation [1] of oscillations in the RHEED intensity
during the epitaxial growth of GaAs, offered a new tool to control thin film
growth with atomic layer precision. In the last few years several research
groups have been able to use this powerful diagnostic technique, in situ, in
combination with Pulsed Laser Deposition (PLD), i.e. the Laser MBE
technique, to obtain the two-dimensional (2D) growth of artificial materials
otherwise difficult or impossible to synthesise. The advantage of the Laser
MBE technique is due to the possibility to monitor the RHEED intensity
oscillations during the growth process, enabling very precise control of the
layer by layer epitaxial growth (phase locked growth) thus reducing as much
as possible occasional fluctuations in the thickness of each constituent
layer. Presently the Pulsed Laser Deposition technique is used to grow a
wide variety of thin oxide films, most of them with a complex structure:
high $T_{c}$ superconductors [2, 3], ferroelectrics [4, 5], piezoelectrics
[6], colossal magnetoresistance oxides [7, 8], electro-optics materials [9].
With the possibilities offered by the utilisation, in situ, of the RHEED
diagnostic, it is possible to predict the realisation of functional
materials of very good quality, with very interesting applications in
several fields [10]. The first observation of RHEED intensity oscillations
on oxide superlattices was carried out by A.Gupta et al. [11] on$%
CaCuO_{2}/SrCuO_{2}$ using during the deposition additional atomic oxygen
source. Recently has been reported the observation of RHEED intensity
oscillations on $(BaCuO_{2})_{2}/(SrCuO_{2})_{2}$ not superconducting
superlattices [12]

In this article we report our results on the hetero-epitaxial growth by
laser MBE and on the RHEED intensity oscillations of the $BaCuO_{x}/CaCuO_{2}
$ artificial structures. As far as we know this is the first time that $%
BaCuO_{x}/CaCuO_{2}$ films have been deposited by Laser MBE, monitoring the
RHEED intensity oscillations. The structure of this superlattice is composed
by stacking, along the c axis, the $BaCuO_{x}$ compound and the
Infinite-Layer (IL) $CaCuO_{2}$ compound [13]. The IL structure consists of
an infinite stack of $CuO_{2}$ planes separated by an alkaline-earth-metal
ion ($Ca$ or $Sr$). It has been shown that such superlattices, when grown by
conventional PLD at high molecular oxygen pressure ($\simeq 9\times 10^{-1}$ 
$mbar$) are superconducting, with a maximum $T_{c}$of $80$ $K$. Furthermore
in Ref. 14 it was pointed out that structural disorder strongly affects the
superconducting properties of these superlattices. Under this respect the
laser MBE technique could allow a better control of the interface disorder.

The depositions were performed using an excimer laser charged with KrF,
generating 248 nm wavelength pulses of 25 ns width with 1Hz repetition rate.
The laser beam, with an energy of 60 mJ per pulse, was focused in a high
vacuum chamber on to the computer controlled multi-target rotating system.
Substrates used for deposition were $(100)$ $SrTiO_{3}$ single crystal,
placed at a distance of about 70 mm from the target on a heated holder.
Before growth, substrates were chemically etched in a buffered solution of $%
NH_{4}F-HF$ ($pH=4.6$) for 8 minutes. This etching treatment leaves a
terminating substrate layer of $TiO_{2}$ and decreases the surface roughness
[15].

Before starting the superlattice deposition, a few monolayers of $SrTiO_{3}$
were deposited, to have an optimal 2D starting deposition surface. The
incident RHEED electron beam was parallel to the [100] substrate direction.
The homoepitaxial $SrTiO_{3}$ deposition was performed under a molecular
oxygen pressure of $10^{-5}mbar$ and with the substrate temperature at $640%
{{}^\circ}%
C$. Four monolayers of $SrTiO_{3}$ were deposited monitoring four RHEED
intensity oscillations of the specular spot. The $SrTiO_{3}$ deposition was
stopped when the specular spot intensity reached the fourth maximum, after
about 200 laser pulses (inset Fig. 1).

To grow the $BaCuO_{x}/CaCuO_{2}$ superlattices, $BaCuO_{x}$ and $CaCuO_{2}$
targets, prepared by solid-state reactions were used. The deposition
temperature was decreased to $500%
{{}^\circ}%
C$ and the molecular oxygen pressure increased to $10^{-4}mbar.$ The
deposition was started with the $BaCuO_{x}$ layer. In these conditions, the
growth rates of $BaCuO_{x}$ and $CaCuO_{2}$ were calibrated, monitoring the
RHEED intensity oscillations of the specular spot. The layer by layer
deposition of $BaCuO_{x}/CaCuO_{2}$ superlattices was carried out stacking
in sequence $m$ $BaCuO_{x}$ unit layers and $n$ $CaCuO_{2}$ unit layers,
(the so called mxn superstructure, $[(BaCuO_{x})_{m}/(CaCuO_{2})_{n}]_{S}$,
where $S$ represents the total number of deposition cycles). The
bi-dimensionality of the growth process is proved by the RHEED pattern
which, at the end of the growth, shows typical 2D features (Fig. 2).

Monitoring the RHEED intensity oscillations, the laser pulses on the two
targets were adjusted to obtain the $BaCuO_{x}/CaCuO_{2}$ superlattice. In
Fig. 1 the RHEED intensity oscillations of the specular streak are shown
during four of the overall 15 cycles of the $(BaCuO_{x})_{2}/(CaCuO_{2})_{2}$
superlattice deposition. The time evolution of RHEED intensity in a single
cycle remains unchanged throughout the superlattice growth. Its major
feature is the sizeable variation of intensity when the growth is switched
from the $BaCuO_{x}$ oxide to the $CaCuO_{2}$ oxide. This effect can be
possibly ascribed to two causes. The first cause is the electron scattering
factors of atoms in the two layers, which differ in their absolute
magnitude, in the phase shift upon scattering and phase shift due to the
height difference. The second cause is the difference in the average surface
quality between the $BaCuO_{x}$ and $CaCuO_{2}$ layer. In addition to the
layer oscillations caused by the composition of the upmost layer, weaker
oscillations can be clearly recognised, in the time evolution pattern of the
RHEED intensity. During the $CaCuO_{2}$ deposition it is possible to note
two weak RHEED intensity oscillations corresponding to the growth of a 2D
layer with a thickness of two unit cells. During the $BaCuO_{x}$ deposition
there was a first faint RHEED intensity oscillation followed by a more
evident one indicating, also in this case, the growth of a 2D layer with a
thickness equal to two $BaCuO_{x}$ cells. The clear asymmetry between the
first and the second oscillation during the deposition of the BaCuOx layer
gives qualitative support to the results of a recent EXAFS study on these
superlattices. In this study it was shown that the structural unit of the $%
BaCuOx$ layer consists of four atomic planes$(CuO_{2}/BaO/CuO_{x}/BaO)$
rather than two identical IL blocks $(CuO_{2}/Ba/CuO_{2}/Ba)$ [16]. Such
RHEED intensity oscillations were observed during all of the 15 cycles. From
the period of the RHEED intensity oscillations, a growth rate of 20 laser
pulses per unit $BaCuO_{x}$ layer and 50 laser pulses per unit $CaCuO_{2}$
layer, was estimated. In Fig. 3 the $\theta -2\theta $ x-ray diffraction
(XRD) spectrum of the same superlattice is shown. 

From the angular distance between the first order satellite peaks (SL$_{-1}$
and SL$_{+1}$) it is possible to obtain the period $\Lambda $ of the
superlattice, $\Lambda =\lambda /(sin\theta _{+1}-sin\theta _{-1})$, where $%
\theta _{+1}$ and $\theta _{-1}$ represent the angular positions of the
first order satellite peaks and $\lambda $ represents the x-ray wavelength.
The average lattice parameter, $\overline{c}$ , can be estimated from the
angular positions of the zero$^{th}$ order SL$_{0}$ (00l) peak, $\theta _{0}$%
,$\overline{c}=\lambda l/2sin\theta _{0}$.

The total number of layers composing the superlattice, $N=m+n$, can then be
calculated, $N=\Lambda /\overline{c}$ . From this simple analysis of the
spectrum, it was calculated that $\Lambda =16.25$ $\AA $ and $N=4.25$, only
slightly greater than the value aimed for ($N=4$). The full width at half
maximum (FWHM) of the rocking curve for the SL$_{-1}$(001) peak of the grown
film is about $0.07%
{{}^\circ}%
$, approximately the same as the (002) $SrTiO_{3}$ FWHM peak. 

A more accurate analysis of the x-ray spectrum was performed using the
procedure described in Ref. [17]. In this reference, numerical simulations
of x-ray spectra were carried out following a kinematical approach using a
simplified model structure. In this approach a two dimensional layer by
layer growth is considered. In such a case, if $m$ and $n$ remain perfectly
constant during the growth of the superlattice, but not corresponding to an
integer number of unit layers, the composition of the mixed layer, formed at
the interface, is always well defined and varies in a controlled fashion
throughout the film thickness (controlled disorder) [18]. The model also
assumes that mixed composition layers can be corrugated to adjust the
internal stresses due to the large mismatch between the constituent oxides.
An additional random disorder is added to take into account the experimental
dispersion in the amount of deposited material in each iteration. Namely, in
each deposition cycle, m and n take random values that follow a Gaussian
distribution with dispersion s around the mean values $<n_{BaCuO_{x}}>$ and $%
<n_{CaCuO_{2}}>$. 

The simulated spectrum (full line) is shown in Fig. 4 together with the
experimental data (empty dots). In order to simulate the experiment the $%
Ca-Ca$ distance was fixed at $3.20$ $\AA $, in accordance with the $c$
lattice parameter of the IL $CaCuO_{2}$ phase [13]. On the other hand, due
to the mobility of oxygen ions in the $BaCuO_{x}$ block, $Ba-Ba$ distances
can vary depending on growth conditions ($O_{2}$ pressure and growth
temperature). From the fit it has been concluded that the $Ba-Ba$ distance
in this case was about $4.56$ \AA\ (slightly greater than the value found in
Ref. 19). The composition of the superlattice was shown to be $%
[(BaCuO_{x})_{1.95}/(CaCuO_{2})_{2.3}]$.

A further important result is that the random disorder parameter s is more
than one order of magnitude smaller, relative to the one used to simulate
similar structures grown without in-situ RHEED diagnostic. In the present
case, the peak broadening, arises mainly from the small film thickness and
the controlled disorder, and in practice s can be considered negligible.

The resistivity versus temperature behavior of the $BaCuO_{x}/CaCuO_{2}$
superlattices was measured by the standard four probe DC technique. It was
found that the resistivity values increase at low temperature following
roughly a Variable Range Hopping mechanism (Fig. 5a). Such a result was
expected due to the low oxygen growth pressure ($\simeq 10^{-4}mbar$). For
comparison in Fig. 5c) is reported the $\rho (T)$ curve for a
superconducting $(BaCuO_{x})_{2}/(CaCuO_{2})_{2}$ superlattice grown without
in situ RHEED diagnostic at higher oxygen pressure ($9\times 10^{-1}mbar$)
and for the pressures equal or lower than $5\times 10^{-2}mbar$ it was not
possible to observe any superconducting transition up to $20$ $K$ (Fig 5 b).

Clearly in order to take advantage from the very good structural quality of
our superlattices, the degree of oxidation during the growth, has to be
still strongly increased.

In conclusion, crystalline $c-$axis oriented $(BaCuO_{x})_{2}/(CaCuO_{2})_{2}
$ superlattice thin films have been successfully grown on (001) $SrTiO_{3}$
substrate by Laser MBE and the growth process of the films investigated by
RHEED. It was shown that RHEED intensity oscillations, even for
superlattices of such complex oxides, can be used for a qualitative phase
locking of the growth. A detailed a posteriori x-ray investigation also
demonstrates that $(BaCuO_{x})_{2}/(CaCuO_{2})_{2}$ superlattices grown by
Laser MBE are only affected by a small amount of controlled disorder, due to
the non exact integer number of unit layers deposited in each iteration,
while the random disorder component is negligible. 

\bigskip 

\bigskip {\Large REFERENCES}

1) J. J. Harris, B.A. Joyce and P. J. Dobson,. Surf. Sci. Lett. 103, L90
(1981).

2) G. Balestrino, S. Martellucci, P. G. Medaglia, A. Paoletti, G.
Petrocelli, Physica C 302, 78 (1998).

3) A. Crisan, G. Balestrino, S. Lavanga, P. G. Medaglia, E. Milani, Physica
C 313, 70 (1999).

4) H. M. Christen, E. D. Specht, D. P. Norton, M. F. Chrisholm, L. A.
Boatner, Appl. Phys. Lett. 72, 2535 (1998).

5) M. Joseph, H. Tabata, T. Kawai, Appl. Phys. Lett. 74, 2534 (1999).

6) M. Tyunina, J. Levoska, S. Leppavuori, Journ. of Appl. Phys. 83, 5489
(1998).

7) M. Izumi, Y. Konishi, T. Nishihara, S. Hayashi, M. Shinohara, M.
Kawasaki, Y. Tokura, Appl. Phys. Lett. 73, 2497 (1998).

8) K. Ghosh, S. B. Ogale, S. P. Pai, M. Robson, E. Li, I. Jin, Z. Dong, R. L
Greene, R. Ramesh, T. Venkatesan, M. Johnson, Appl. Phys. Lett. 73, 689
(1998).

9) Y. Shibata, K. Kaya, K. Akashi, M. Kanai, T. Kawai, S. Kawai, Appl. Phys.
Lett. 61, 1000 (1992).

10) L. R. Tagirov, Phys. Rev. Lett. 83, 2058 (1999).

11) A.Gupta, T. Shaw, M.Y. Chern, B.W. Hussey, A.M. Guloy and B.A. Scott.,
Journ. Sol. State Chem. 111, 1 (1994).

12) G. Koster, G.J.H.M. Rijnders, D.H.A. Blank and H. Rogalla, Appl. Phys.
Lett. 74, 3729, (1999)

13) G. Balestrino, R. Desfeux, S. Martellucci, A. Paoletti, G. Petrocelli,
A. Tebano, B. Mercey, M. Hervieu, J. Mater. Chem. 5, 1879 (1995).

14) G. Balestrino, A.Crisan, S. Lavanga, P.G. Medaglia, Petrocelli, A.A.
Varlamov, Phys. Rev. B 60, 10505 (1999).

15) M. Kawasaky, K. Takahashi, T. Maeda, R. Tsuchiya, M. Shinohara, O.
Lshiyama, T. Yonezawa, M. Yoshimoto, and H. Koinuma, Science 266, 1400
(1989).

16) S. Colonna, F. Arciprete, A. Balzarotti, G. Balestrino, P. G. Medaglia,
G. Petrocelli, Physica C 334, 64 (2000).

17) G. Balestrino, G. Pasquini, A. Tebano, Phys. Rev. B 62, 1421, (2000).

18) I. K. Shuller, M. Grimsditch, F. Chambers, G. Devane, H. Vanderstraeten,
D. Neerinck, J. P. Loquet and Y. Bruynseraede, Phys. Rev. Lett. 65, 1235
(1990).

19) F. Arciprete, G.Balestrino, S. Martellucci, P.G. Medaglia, A. Paoletti,
G. Petrocelli, Appl. Phys. Lett. 71, 959 (1997).

\bigskip 

\bigskip 

\bigskip 

\bigskip 

\bigskip 

\bigskip 

\bigskip 

\bigskip 

\bigskip 

\bigskip 

\bigskip 

\bigskip 

\bigskip 

\bigskip 

\bigskip 

\bigskip 

\bigskip 

\bigskip 

\bigskip 

\bigskip 

\bigskip 

\bigskip 

\bigskip 

\bigskip 

\bigskip 

$FIGURE\;CAPTIONS$

Fig. 1) RHEED intensity oscillations during the hetero-epitaxial growth of
the superlattice $BaCuO_{x}/CaCuO_{2}$ as a function of the laser pulses. In
the inset are shown the four RHEED intensity oscillations monitored during
the homo-epitaxial deposition of $SrTiO_{3}$.

Fig. 2) RHEED pattern at the end of the deposition process of a $%
[(BaCuO_{x})_{2}/(CaCuO_{2})_{2}]_{15}$ artificial superlattice.

Fig. 3) XRD experimental spectrum of the $BaCuO_{x}/CaCuO_{2}$ superlattice.

Fig.4) XRD experimental spectrum of the $(BaCuO_{x})_{2}/(CaCuO_{2})_{2}$
superlattice (open circles) compared with the simulated spectra (continuous
line). 

Fig. 5) Resitivity vs. Temperature measurements for the $BaCuO_{x}/CaCuO_{2}$
superlattice grown in different conditions: (a) $10^{-4}mbar$ of oxygen
pressure (Laser MBE), (b) $5\times 10^{-2}mbar$ of oxygen pressure (PLD),
(c) $9\times 10^{-1}mbar$ of oxygen pressure (PLD).

\end{document}